\documentclass[a4paper,11pt]{article}
\usepackage[utf8]{inputenc}
\usepackage[T1]{fontenc}
\usepackage{authblk}
\usepackage{authblk}
\usepackage{amsthm,amsmath,amssymb,graphicx,color,bbm}
\usepackage[linesnumbered,ruled]{algorithm2e}
\usepackage{caption}
\usepackage{subcaption}
\usepackage[tableposition=top]{caption}

\newcommand{\bc}{\begin{center}}
\newcommand{\ec}{\end{center}}
\newcommand{\be}{\begin{equation}}
\newcommand{\ee}{\end{equation}}
\newcommand{\bq}{\begin{eqnarray}}
\newcommand{\eq}{\end{eqnarray}}
\newcommand{\nn}{\nonumber}

\newcommand{\ed}{\end{document}}

\newcommand{\bea}{\begin{eqnarray}}
\newcommand{\eea}{\end{eqnarray}}

\def\m1{\mathbbm{1}}

\def\eq{E^{\mathbb{Q}}_0}

\title{Using Machine Learning to Predict Realized Variance}
\author[1]{Peter Carr\thanks{petercarr@nyu.edu}}
\author[2]{Liuren Wu\thanks{liuren.wu@baruch.cuny.edu}}
\author[1]{Zhibai Zhang\thanks{z.zhibai@gmail.com, corresponding author}}

\affil[1]{Department of Finance and Risk Engineering, NYU Tandon School of Engineering}
\affil[2]{Zicklin School of Business, Baruch College, City University of New York}

\begin{document}
  \maketitle

\begin{abstract}

In this paper we formulate a regression problem to predict realized volatility by  using option price data and enhance VIX-styled volatility indices' predictability and liquidity. We test algorithms including regularized regression and machine learning  methods such as Feedforward Neural Networks (FNN) on S\&P 500 Index and its option data. By conducting a time series validation we find that both Ridge regression and FNN  can improve volatility indexing with higher prediction performance and fewer options required. The best approach found is to predict the difference between the realized volatility and the VIX-styled index's prediction rather than to predict the realized volatility directly, representing a successful  combination of human learning and machine learning.  We also discuss suitability of different regression algorithms for volatility indexing and applications of our findings.

\end{abstract}

\newpage

\section{Introduction}

Forecasting realized volatility is an essential part for option pricing, trading, portfolio construction and risk management. In option pricing, the volatility dynamics is usually described by models with very few parameters in the risk neutral measure. For instance, in the Black-Scholes model  \cite{b76} a security's volatility is simply set to be a constant. This is apparently oversimplified, compared to the behavior of the financial market. Nonetheless, these simple parametric models are useful as they gives one a way to build a projection of volatility in a controllable fashion. In the case of Black-Scholes formula, instead of pricing an option, it is more often used to extract the volatility implied by an option's market price. This measure, referred to as the implied volatility, reflects option traders' view of the realized volatility of the security. For this reason, implied volatility is often regarded as a benchmark  for realized volatility forecast. 

Since there are numerous options written on an optionable security and each of them has  different implied volatility,  selecting the most proper options for realized volatility forecast is non-trivial.  The task of selecting options and aggregating them into a single volatility forecast is called volatility indexing.  The most well known index in this category is the CBOE Volatility Index (VIX), which is composed of options with close to 30-day maturity on S\&P 500 index (SPX). It is one of the best estimators of 30-day future realized volatility of S\&P 500 Index. There is an increasing number of volatility indices on various equity indices and major instruments in fixed-income, currency and commodity in recent years. In VIX-styled indices, the option weights are determined by a variance swap pricing formula such that the index's squared payoff replicates the underlying's variance  in the risk neutral measure. Albeit its popularity, the VIX-styled indexing possesses some caveats. First, as the weights are set in risk-neutral measure, it is not certain if the weighting scheme has the optimal forecastability, especially out-of-sample (OOS) in the market measure.   Additionally, it involves a large number of out-of-the money options, which have ascending illiquidity. That makes it expensive for the volatility index sellers to hedge, as they would need to hold many thinly traded options. 

In this paper we investigate these issues by asking the following questions: (1) is the current VIX-styled weighting scheme optimal in predicting realized variance? (2) how can one use prediction models such as machine learning to improve volatility indexing? (3) is it feasible to achieve good predictability with fewer options? To address these questions, we formulate a regression problem to predict realized volatility by predominately using option price data. We employ algorithms including linear and regularized regression, and machine learning  techniques such as Feedforward Neural Networks (FNN), and impose constraints on model selection to make sure the prediction can be replicated by an volatility index. We test the algorithms by building a time series validation scheme on SPX and its option data. We  discover that by combing the prediction model and the VIX-styled weighting scheme, one can achieve a volatility index that has improved predictability and liquidity. The best performing approach found is to use Ridge and  machine learning regression to forecast the deviation between the realized volatility and the VIX-styled index's prediction, which is commonly referred to as the variance risk premium. Therefore, it represents a successful combination of human learning and machine learning. This is encouraging as intuitively it indicates that cooperation of human and machine actually yields better results than each of them individually in our volatility indexing framework. We also discuss suitability of different regression algorithms for volatility indexing. As we will show, the tradability condition in fact imposes a strong constraint on algorithm selection, which most models do not satisfy except for piece-wise linear ones such as FNN with a ReLU activation function. 

This paper joins a large number of literature on volatility forecasting. In the past, there has been great progress on time series based models such as ARCH/GARCH \cite{engle82, bollerslev86} and HAR \cite{corsi09}. In these models it has been shown that historic volatility measures exhibit predictability at future realized volatility. To this end, we also experiment with historic volatility features as alternative tests to the main model and we find that their contribution is limited in our framework. There has also been abundant study on the predictability of option price and implied volatility at realized volatility \cite{F08, torben07, fleming98, busch11}. More recently, application of machine learning techniques in volatility forecasting have emerged \cite{chuong18, shaikh04}. To our knowledge, this paper is the first attempt to apply regularized regression and machine learning to improve volatility indices' forecastability.   

The rest of the paper is organized as follows: In section 2 we review the details of VIX-styled volatility indices' weighting scheme. In section 3 we show how to process option data to formulate a realized volatility prediction problem. We also discuss model selection and evaluation. This is followed by section 4, where we present the main result on prediction performance. Section 5 concludes and presents directions for future research.   

\section{Volatility prediction and indexing}

We consider volatility of daily returns. The realized return variance between $t$ and $t+T$ is 
\be
Var_{t}^T=\frac 1T \sum_{i=1}^T(r_{t+i}-\bar{r})^2,
\ee
where $r_j=\frac{p_j-p_{j-1}}{p_j}$ is the security's daily return at time $j$ and $\bar{r}$ is the average return between $t+1$ and $t+T$. Sometimes the mean return $\bar{r}$ is omitted as it is close to zero in most cases. Volatility is the square root of the variance and it is more often quoted in the derivative markets. In this paper, we use the terms volatility and variance interchangeably. Volatility distribution tends to be skewed and fat tailed, with high volatility events sparsely distributed among low volatility periods across time. High volatility events also tend to appear sequentially, a phenomenon   referred to as volatility clustering. This bring the difficulty of volatility forecast: it is common to overestimate during low volatility periods and underestimate during high volatility.

\subsection{The construction of VIX}
The CBOE Volatility Index (VIX) is one of the most broadly used volatility measure for both SPX and the entire market. The  value of VIX divided by $100$ is often used as an estimator of SPX's 30-day realized volatility. VIX constituents a family of European options written on SPX Index. On each day, one selects `near-term' options whose maturity $T_1$ is the closest to yet less than 30 days, and `next-term' options whose maturity $T_2$ is the closest to yet greater than 30 days. Secondly, for each term, one selects out-of-the-money call and put options with strikes $K$ centered around the at-the-money strike price $K_0$. For both terms, all strikes are included until two consecutive strikes are missing a quote. After the options are selected, one computes the implied variance for each term:
\be
\sigma_1^2=\frac{2}{T_1}\sum_h \frac{\Delta K}{K_h^2}e^{R_1T_1}O(K_h,T_1)-\frac 1{T_1}(\frac{F}{K_0}-1)^2\,,\nn
\ee
\be
\sigma_2^2=\frac{2}{T_2}\sum_h \frac{\Delta K}{K_h^2}e^{R_2T_2}O(K_h,T_2)-\frac 1{T_2}(\frac{F}{K_0}-1)^2\,, \label{sigmas}
\ee
where $\Delta K=K_{h+1}-K_h$ is the spacing in strike prices, $R_i$ is the interest rate for each term, $F$ is the forward index value. The second term on the r-h-s in each equation is considerably smaller than the first term. Finally VIX is computed as the square root of a weighted sum of the r-h-s in the above equations up to scaling by 100:
\be
\frac{\rm{VIX}}{100}=\sqrt{T_1 \sigma_1^2 \left(\frac{T_2-T_{30}}{T_2-T_1}\right)+T_2 \sigma_2^2 \left(\frac{T_{30}-T_{1}}{T_2-T_1}\right)}\label{vix}\,.
\ee
Note that selecting the two terms and linearly interpolating them is to ensure the effective time-to-maturity is exactly 30 days. When there is options with 30 day time-to-maturity, one can simply compute the variance term in Eqn (\ref{sigmas}) on them. For more details, see \cite{vixwhite}.

Roughly speaking, $\rm{VIX}^2$ is a portfolio of out-of-the-money (OTM) options with weights inversely proportional to the squared strike prices:
\be
{\rm{VIX}}^2\sim \sum \frac{2\Delta K}{K_h^2}O(K_h,T)\,.\nn
\ee
The $\frac{2\Delta K}{K^2}$ weighting scheme comes from the variance swap pricing formula  \cite{carr04}. The weights are determined such that the portfolio's payoff perfectly replicates the variance of the underlying in risk neutral measure. However, when it comes to forecasting realized volatility in the market measure, there is no guarantee that this weighting scheme's prediction is optimal. Furthermore, the option selection criteria normally produce a large number of options. For instance, in the example in \cite{vixwhite}, with the spot price at $\$1960$, the lowest put strike is at $\$ 1370$ while the highest call strike is at $\$ 2125$, which contains 149 options in total. Holding that many OTM options is extremely costy as the liquidity is very low for deep OTM options in general. This makes hedging challenging for VIX sellers.

\section{Formulate a machine learning regression problem}
In this section we set up a new way to determine weights for volatility indices using regularized linear and machine learning regression. A supervised ML algorithm can be summarized as a function approximation problem aimed to find a function $f(\cdot)$ such that
\bea
y&=&\hat{f}(\vec{x})\,,\nn\\
\hat{f}&=&\arg \min\left\{{\rm{Err}}(f(\vec{x}),y)\right\}\label{ml}
\eea
where ${\rm{Err}}(f(\vec{x}),y)$ a pre-defined object function defined on samples of $(y,\vec{x})$. For many ML algorithms, $f(\cdot)$ is either semi-parametric (with a large number of parameters) or non-parametric (can only be carried out operationally and does not have a closed form expression). $y$ is usually referred to as target value and entries in $\vec{x}$ are referred to as features. In the following, we formulate a ML regression problem by constructing target value and features from realized volatility and option price respectively. 

\subsection{Two regression approaches}
Our first approach is to directly model future realized variance as a function of option price. Mathematically, this is
\bea
Var_t^T=f\left(\left\{ O_t(K_i, T'_j) \right\} \right) + \epsilon_t\,,\label{reg1}
\eea
where $\{ O_t(K_i, T'_j) \} $ is all the options across selected strikes and tenors and $\epsilon_t$ is a zero-mean noise term. As mentioned above, the function $f$ will be determined by fitting the algorithm on the training data set. We call this approach Regression I.

A ML based function approximation such as Eqn (\ref{reg1}) can be useful in estimating functions with high non-linearity and non-parametricity. This makes Eqn (\ref{reg1}) desirable as realized variance is expected to be strongly non-linear. However, the large number of parameters and complex optimization involved in ML algorithms may also be problematic, as they can lead to overfitting due to outliers in the training set (e.g. extremely high volatility events). Additionally, as VIX-styled weighting scheme already contains some predictive power, in principal we would like to incorporate it with the regression approach too. This lead to our second regression approach:
 \bea 
 Var_t^T=VIX^*\left(\left\{ O_t(K_i, T'_j) \right\} \right) ^2+f\left(\left\{ O_t(K_i, T'_j) \right\} \right) + \epsilon_t\,,\label{reg2}
\eea
where $VIX^*$ is the synthetic VIX index using the selected options.The only difference between the two is that $VIX^*$ contains a fixed number of options throughout time.  We call approach Eqn (\ref{reg2}) Regression II.  

Note that if one switches the ${VIX^*}^2$ to the l-h-s, then it is equivalent to regress a variance risk premium (VRP) proxy, which is
\bea
VRP_t = {VIX^*}^2_t - Var_t\,.
\eea
So this approach is to train the algorithm to forecast $VRP_t$ and combine  it with $VIX^*$ to forecast realized volatility.  Though Eqn (\ref{reg2}) is a special case of Eqn (\ref{reg1}), the two approaches in general produce different results in non-linear cases. Regressing the difference between $Var_t^T$ and ${VIX^*}^2$ rather than directly regressing $Var_t^T$ can be considered as  a normalization procedure. In nonlinear models,  proper normalization can often substantially improve an algorithm's fitting and performance. We will show that Regression II indeed outperforms Regression I.

\subsection{Data processing and feature generation}\label{feagen} 
We use daily data of  options written on SPX with a time span from 1996 to 2016. The option market data is obtained from OptionMetrics. On each trading day, the data set contains midquotes of options with multiple maturities and strikes. Furthermore, interest rate and SPX spot and  forward price data are also used. 

Generating features from option price time series  proves to be a nontrivial task. There are two aspects of this data set that raises problems for a machine learning formulation. First, the actively traded options are those whose strikes are centered around the spot, which varies all the time. As a result, the set of OTM options needs to be re-selected everyday. Moreover, the maturities decrease until expiry. In general, most machine learning algorithms require the features to be stationary. Since our goal is to use option price to forecast the realized volatility with a fixed horizon, it is also important to have the options' maturities in sync with the forecast horizon. Fortunately, the construction of VIX provides a solution to the varying maturity issue. Namely, we can linearly interpolate the option with the closest maturities to the forecast horizon as in Eqn (\ref{vix}). Since regression problems require a fixed number of features, we need to select a constant number OTM options with strikes centered around the spot price. Put together, we generate raw option price features as follows:
\be
\tilde{O}_t\left(K_h,T\right)=O_t\left(K_h,T_1\right)\left(\frac{T_2-T}{T_2-T_1}\right)+O_t\left(K_h,T_2\right)\left(\frac{T-T_1}{T_2-T_1}\right)\,,\label{fea}
\ee 
where $T$ is the forecast horizon (in the case of VIX this is 30 day), $T_1$ and $T_2$ are the two closest existing maturities to $T$ of all available options at day $t$. On each day, we select $N=2n+1$ of strikes $K_h$ centered around $K_0$:
\be
\{K_{-n}, K_{-n+1},...,K_{-1},K_0,K_1,...,K_{n-1},K_{n}\}\,,\label{strikes}
\ee
where $\Delta K=K_{i}-K_{i-1}$ (for SPX  options, $\Delta K=\$5$). It needs to be understood that the at-the-money strike $K_0$ changes every day following the spot price $S_t$. Therefore, the strike set Eqn (\ref{strikes}) is determined on a daily basis. Notice that as $n$ becomes larger, the corresponding strike $K_{n}$ is more out-of-the-money and the option tends to be less frequently traded. If for a specific $K_h$, the option does not have a quote, we linearly interpolate the midquote using midquotes of the options with the two closest strikes to $K_h$. This way, Eqn (\ref{fea}) gives rise to a consistent feature generation once the number of features $N$ and forecast horizon $T$ are chosen for an option time series.

In machine learning for most algorithms it is important to normalize features to meet stationarity conditions. For each feature, a common practice is to subtract its  sample mean and divide it by its  sample standard deviation. We apply this technique and normalize the features using the training sets. In addition, there is subtlety related to financial time series, which is that a security's price is a non-stationary process as otherwise there is arbitrage. Therefore, option price features in Eqn (\ref{fea}) contain this specific nonstationarity that cannot be removed by the standard machine learning normalization procedure. A rationale is that, an option on SPX worth \$5 in 2010 is not the same as an option worth the same thing in 2015, as the level of  SPX has grown considerably over that period. For this reason, we apply the following pre-processing before the standard ML normalization to the option price features in Eqn (\ref{fea}):
\be
\bar{O}_t\left(K_h,T\right)=\frac{\tilde{O}_t\left(K_h,T\right)}{K_h^2}\,.\label{fea1}
\ee

In addition to the main option price features, we also experiment with  returns-based idiosyncratic features of SPX as alternative tests. These include realized returns and variance with different lookback windows.
\begin{itemize}
\item Realized returns, $r_{t,l} =\frac{p_t-p_{t-l}}{p_{t-l}}$ with $ l \in \{1, 5,15,30,60,90\}$
\item Realized variance, $var_{t,l}=\frac 1l \sum_{i=1}^l(r_{t-i}-\bar{r})^2$, with $ l \in \{15,30,60,90\}$   
\end{itemize}
We only consider these features in combination with the option features, but not their predictability individually. This is mainly because that once these features are added, the prediction is no longer tradable as they cannot be replicated by any portfolio.  

\subsection{Algorithms}
One advantage of machine learning is that the algorithm can be highly adaptive. In the form of Eqn (\ref{ml}), this  means that $\hat{f}$ usually does not have a close form expression. Since we want to be focused on an indexing problem, this may cause trouble. In both of the two regression approaches Eqn (\ref{reg1},\ref{reg2}), if we require that the forecast can be replicated by an option portfolio, we need  $f(\cdot)$ to be at least a piece-wise linear function. This is actually a strong constraint that renders many ML regression methods not suitable. For instance, for K-nearest neighbors regression the forecast is made on the sub-sample mean of the training set and is clearly not piece-wise linear. In that case, obviously one cannot build an option portfolio whose payout is the forecast. 

In this paper, we consider four regression algorithms: linear regression, ridge regression, feedforward neural network regression with ReLU activation function, and random forest regression. We start off with linear regression for its simplicity. Compared with ML algorithms, linear regression has the advantage that it does not involve tunable parameters which makes it the least prone to overfitting. However, linear regression also has a few shortcomings, such as its lacking of non-linearity. Another potential issue is that, since our main feature set is a basket of options that have strong correlation, it may cause trouble for the ordinary least square fitting. For this reason, we consider ridge regression, which has the same functional form as linear regression but the error function is $L2$ regularized.

Feedforward neural network (FNN) is one of the simplest neural network models. The graphic representation of a neural network is composed of a number of layers, including an input layer corresponding to the features and an output layer corresponding the labels. The rest is referred to as hidden layers. Each layer contains several neurons and different layers are connected by a certain topology. Mathematically, both neurons and connections between neurons correspond to variables of the prediction functions. A feedforward neural network with one hidden layers has the following functional form
\bea
y(\pmb x, \pmb w)=O\left( \sum_j w^{(2)}_{j} \cdot  h\left(\sum_i w^{(1)}_{ji} x_i  +w^{(1)}_{j0}\right) +w^{(2)}_{0} \right)\,,\label{fnn}
\eea
where $O$ and $H$ are activation functions on the output and hidden layers, $w^{p}_{ji}$ is connection weight that connects the $i$-th neuron on the $(p-1)$-th layer to the $j$-th neuron on the $p$-th layer. We select rectified linear unit (ReLU) as the activation function for both the hidden layer and output layer:
\be
ReLU(x)=max\{0, x\}\,.
\ee
With a ReLU FNN, $y$ is piecewise linear:
 \be
y(\pmb x, \pmb w)=\sum_{i,j} \m1_{(2),j}  w^{(2)}_{j}  w^{(1)}_{ji} x_i + const\,.
\ee
where $\m1_{(p),k}$ is an indicator function that corresponds to the ReLU activation function on the $k$-th neuron on the $p$-th layer. Now it is evident that ReLU FNN is piece-wise linear and thus suitable for our tradability constraint. For model simplicity, we construct FNN with only one hidden layer, whose number of neurons is the same as the input layer. 

On the contrary, some ML algorithms are not piece-wise linear. Random forest is a very popular algorithm in this category. Random forest is an ensemble algorithm that consists of multiple base algorithms called decision trees. For decision tree regression, the algorithm iteratively splits the input data on the features such that the information gain from splitting a parent sample to children samples is optimized. After the algorithm is trained, each sample (regardless in the training or the test set) can be run through the tree and land on one of the many sub-samples after splitting the original training sample. The prediction is then  given as the mean value of the dependent variable in the specific sub-sample. In random forest regression, multiple decision tree regressions are fitted, each on a randomly sampled training set to increase robustness, and the final prediction is an average of the prediction of all decision trees. So schematically, the forecast from random forest regression  is
\be
y(x)=\frac{1}{N_I} \sum_{i\in I} y_i\,,\label{rf}
\ee
where $I$ is the subset the $(x,y)$ belongs to, $N_I$ is the total number of samples in this subset, and $y_i$ is the value of the dependent variable of the $i$-th sample in the subset. Relating to our setting, this means that when using random forest, the predicted variance is a function of the realized variances of at selected past timestamps, which clearly cannot be the payoff of an option portfolio. More straightforwardly, Eqn (\ref{rf}) is not a piece-wise linear function. Nonetheless, we will keep random forest in the test for predictability comparison. 

\subsection{Validation, evaluation and model calibration}\label{validation}
\begin{figure}
\centering
\includegraphics[height=5cm]{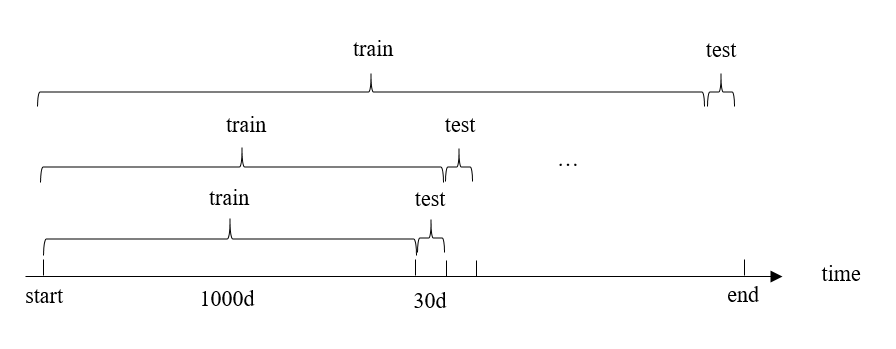}
\caption[E]{\footnotesize An illustration of the rolling validation process.}\label{tvval}
\end{figure}

To evaluate all algorithms and feature combinations, we formulate an out-of-sample (OOS) test on the data set in a time series fashion. To do so, we reserve the first 1000 observations for initial training, and we make OOS prediction on the first 30 observations on the remaining set. We re-train the model after every 30 observations with all the available observations on a rolling basis. Every time the training set is purged to get rid of the observations that has an informational overlap with the test set (see \cite{mldp2018} for similar techniques). Obviously a different combination of the size of each training set and the frequency of model retraining may vary the performance of each model. We do not test other validation settings as this may cause overfitting due to the limited amount of data. A graphic representation of the OOS test is shown in Fig (\ref{tvval}).

For prediction performance metrics, we use OOS $R^2$ 
\be
R^2=1-\frac{\sum_i (y_i-p_i)^2}{\sum_i (y_i-\bar{y})^2},
\ee
where $y_i$ is the actual realized variance for sample $i$, $\bar{y}$ is the mean realized variance of all samples and $p_i$ is the model predicted realized variance for sample $i$. For Regression II (\ref{reg2}), even though the direct prediction is the negative variance risk premium, we convert it back to realized variance by adding  ${{VIX^*}^2}$ when computing the $R^2$. This way the $R^2$ for both regression approaches is comparable. 

Many ML algorithms contain multiple hyperparameters that need to be tuned. A common practice is to optimize a large number of hyperparameter combinations and choose the one that performs the best. However, this makes it very easy to overfit for financial time series data. As mentioned above, for each non-linear regression we only optimize one hyperparameter to minimize potential overfitting:
 \begin{itemize}
\item for Ridge, the $L2$ regularization strength $\lambda \in \{10^{-3}, 10^{-2},10^{-1},1,10^{2},10^{3}, 10^{4}\}$
\item for ReLU FNN, the $L2$ regularization strength $\lambda \in \{10^{-3}, 10^{-2},10^{-1},1,10^{2},10^{3}, 10^{4}\}$
\item for Random Forest, maximum tree depth $\in \{3,5,10, \infty\}$
\end{itemize}
All the algorithms are implemented in Python using the Scikit-learn package \cite{sklearn}.

\section{Main Results}
In this section we present our main finding. First, we show the OOS $R^2$ for both regression methods with all four algorithms. We conduct the experiment with a varying number of consecutive strike prices and two different forecast horizons. The performance is compared across different algorithms. Then we report the results of some alternative tests, including using non-consecutive strike prices and adding historic volatility and returns as extra features. 

\subsection{ $T=30$ days}
The first forecast horizon we test is 30 days, which is the designated horizon for VIX. For each algorithm, we use $2k+1$ consecutive options as features ($k$ OTM put, $k$ OTM call and 1 ATM), with $k \in \{10, 20,30,40\}$. We present the performance for the two regression approaches, and highlight the best performing algorithm with a specific number of options.

\begin{itemize}
\item
 OOS $R^2$ for Regression I, 
\begin{center}
\begin{tabular}{|c |c|c|c|c|c|}
\hline
 \# of options & ${VIX^*}^2$&    Linear  & Ridge & RF & FNN \\
\hline
\hline
21 & 0.169 & 0.313 & \textbf{0.320} & -0.018 & 0.154 \\
41 & 0.313 &  0.191&  \textbf{0.329}& -0.021 & 0.114 \\
61 & \textbf{0.366} &  0.163&  0.339& -0.023 & 0.251 \\
81 & \textbf{0.389} &  0.153&  0.334& -0.026 & 0.339 \\
\hline
\end{tabular}
\end{center}
\item
OOS $R^2$ for Regression II, 
\begin{center}
\begin{tabular}{|c |c|c|c|c|c|}
\hline
 \# of options & ${VIX^*}^2$&    Linear  & Ridge & RF & FNN \\
\hline
\hline
21 & 0.169 &  \textbf{0.313} & 0.320 & 0.227 & 0.228 \\
41 & 0.313 &  0.191&  0.329 & 0.328 & \textbf{0.332} \\
61 & 0.366 &  0.163&  0.371& 0.372 & \textbf{0.373} \\
81 & 0.389 &  0.153&  0.392& 0.392 & \textbf{0.394} \\
\hline
\end{tabular}
\end{center}

\end{itemize}
Combining the two regression methods' results and best performance of each algorithm for a specif number of options, the overall model comparison is shown in Fig (\ref{opt30}). First, it is apparent that Regression II produces greater performance than Regression I, as the OOS $R^2$ for the former is higher than that of the latter for all non-linear algorithms (for $VIX^*$ and linear regression, I and II are equivalent). It is also obvious that including more options enhances  OOS $R^2$, except for linear regression. This is because adding more options whose prices are highly correlated deteriorates linear regression's fitting. Nonetheless, when the number of options is small ($21$), linear regression actually outperforms the others. As the number of options increases, FNN with Regression II becomes the best one, though the difference to other methods is only incremental \footnote{In \cite{gu2018}, the authors report a similar incremental enhancement from ML in the case of asset pricing}. 

\begin{center}
\includegraphics[height=7.5cm]{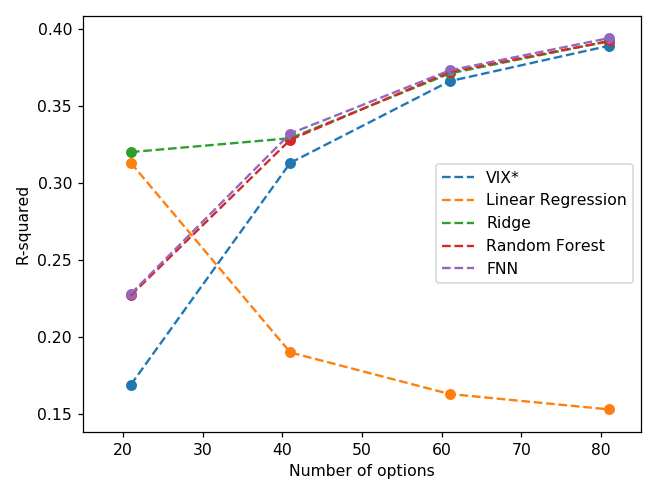}\label{opt30}
\captionof{figure}{$T=30$ days. OOS R-squared for different algorithms, optimal performance combining Regression I and II.}
\end{center}

\subsection{ $T=60$ days} 
Next we conduct a similar test with $T=60$ days. We present the performance for the two regression approaches, and highlight the best performing algorithm with a specific number of options.
\begin{itemize}
\item
 OOS $R^2$ for Regression I, 
\begin{center}
\begin{tabular}{|c |c|c|c|c|c|}
\hline
 \# of options & ${VIX^*}^2$&    Linear  & Ridge & RF & FNN \\
\hline
\hline
21 & \textbf{0.300} &  0.264&  0.266& -0.015 & 0.094 \\
41 & 0.206 &  0.269&  \textbf{0.293}& -0.023 & 0.099 \\
61 & 0.014 &  0.247&  \textbf{0.303}&  -0.022& 0.140 \\
81 & -0.155 &  0.216&  \textbf{0.313} & -0.025& 0.146 \\
\hline
\end{tabular}
\end{center}
\item
OOS $R^2$ for Regression II, 
\begin{center}
\begin{tabular}{|c |c|c|c|c|c|}
\hline
 \# of options & ${VIX^*}^2$&    Linear  & Ridge & RF & FNN \\
\hline
\hline
21 & \textbf{0.300} &  0.264& 0.296 &0.297  & 0.297 \\
41 & 0.206 &  0.269&  0.339 & 0.340 & \textbf{0.342} \\
61 & 0.014 &  0.247& 0.311 & 0.300 & \textbf{0.311} \\
81 & -0.155 & 0.216 & \textbf{0.299} & 0.244  & 0.280 \\
\hline
\end{tabular}
\end{center}

\end{itemize}
Combining the two regression methods' results and best performance of each algorithm for a specif number of options, the overall model comparison is shown in Fig (\ref{opt60}). Interestingly, as the forecast horizon goes up the benefit of including more options diminishes. In most cases, the performance becomes worse after the number of options is above a certain value. For $60$ day horizon the optimal number of options is $41$, with FNN, Ridge and Random Forest being quite close. 

\begin{center}
\includegraphics[height=7.5cm]{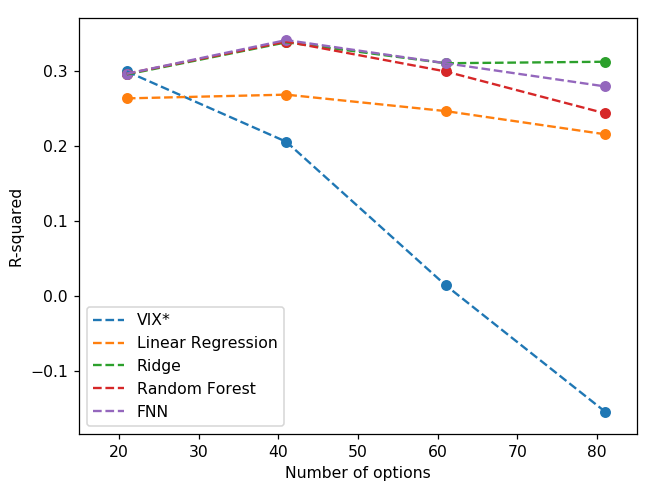}\label{opt60}
\captionof{figure}{$T=60$ days. OOS R-squared for different algorithms, optimal performance combining Regression I and II.}
\end{center}


\subsection{Alternative tests}
To further verify our models' predictability, we run a couple of alternative tests in addition to the results reported above. First, as mentioned in Section (\ref{feagen}), we add returns based features to the main option based features. Since a key of our approach is to maintain both predictability and tradability, it is important to note that these features are not tradable and cannot be included in an index. Nonetheless, we add these features to see how much predictability they may contribute to the main option features. To do this, we re-run the OOS test using Regression II for $T=30$ days, with 41 consecutive options and the 10 realized returns and variance features. The OOS $R^2$'s are: 
\begin{center}
\begin{tabular}{|c |c|c|}
\hline
 algorithms & Options & Options and realized returns/variance    \\
\hline
Linear & 0.191& 0.124\\
Ridge & 0.329&0.335\\
Random Forest & 0.328&0.332\\
FNN & 0.332& 0.335\\
\hline
\end{tabular}
\end{center}
It shows that except for linear regression, including realized returns and variance features can improve the performance, even though the change is incremental. 

Since the options of the same underlying tend to have strong correlations, it is interesting to see if one can increase the spacing in strikes when selecting options. By selecting options with larger distance in their strikes, one gets the benefit of using fewer options, which corresponds to greater liquidity. So far we have only tested the algorithms with consecutive strikes, namely, the spacing is \$5. Here we re-run FNN using regression II with 21 and 41 options and increased spacing of \$10. The results are quite interesting:
  \begin{center}
\begin{tabular}{|c |c|c|}
\hline
\# of options & $\Delta K = 5$ & $\Delta K =10 $  \\
\hline
21 & 0.228& 0.254\\
41 & 0.373&0.268\\
\hline
\end{tabular}
\end{center}
It shows that having consecutive strikes is in fact necessary to maintain the predictability, as when the number of options is $41$ the performances with $\Delta K=5$ and $\Delta K=10$ diverge and the larger spacing set performs less well.

\section{Conclusion}
In this paper we focused on a machine learning based realized variance prediction and indexing problem. Inspired by VIX's predictability to SPX's 30 days realized variance, we have proposed an approach that uses option price as features combined with regression techniques to forecast a securities realized variance. The selection of options in our approach is analogous to that of VIX, with the difference coming from constraints of regression techniques. Based on this we have generated a synthetic volatility index $VIX^*$ and used its squared value to as a benchmark predictor. Then we formulated a regression problem to predict the realized variance and its deviation from    ${VIX^*}^2$. A number of traditional and machine learning regression algorithms, including random forest and FNN have  been tested using a rolling OOS test procedure. 

We have found that (1) when predicting realized variance,  it is best to predict its deviation from the synthetic ${VIX^*}^2$ using a non-linear model. This indicates that there is non-linearity in the volatility process in the p measure; (2) both Ridge and machine learning algorithms can give better prediction performance; (3)  adding more options can increase the predictability of most models. But we note that this also has the drawback of imposing liquidity issues. 

We have also shown that in the task of volatility indexing, there is a constraint on the prediction algorithm selection from tradability. If one expects to have an portfolio whose payoff is the realized variance, the features have to be option price and the regression algorithm is required to be piece-wise linear. In particular, the best performing method, FNN with ReLU satisfies the tradability condition. On the contrary, FNN with other non-linear activation functions and methods such as random forest do not meet this condition.

Before closing, we highlight a few future directions along the line. First it is worth looking into the predictability of other predictors such as macro and cross asset factors. In the framework of machine learning, these are simply additional features. If these features are tradable, namly, they can be replicated by tradable securities, then the method can be generalized to  a broader  volatility index consisting of options and other instruments.  Secondly, it will be useful to  test the framework with higher frequency  data. This way, with potentially much larger data sets, the power of machine learning may be enhanced. It will also be interesting to generalize the Regression II approach in this paper to other forecast problems and investigate whether ML can improve existing parametric models. As shown in this paper,  the human + machine learning approach can outperform each of them individually. It will be intriguing to investigate if this works in other areas. 

\section*{Acknowledgment}
We would like thank Vasant Dhar, Rajesh Krishnamachari, Dacheng Xiu and Haoxiang Zhu for their valuable comments and suggestions.  They
are not responsible for any errors.

\end{document}